\documentclass[fleqn,usenatbib]{mnras}

\usepackage{newtxtext,newtxmath}

\usepackage[T1]{fontenc}
\usepackage{ae,aecompl}

\usepackage{graphicx}	
\usepackage{amsmath}	
\usepackage{multicol}        
\usepackage{bm}		
\usepackage[usenames]{color}
\usepackage{xcolor}
\usepackage[normalem]{ulem}
\definecolor{darkgreen}{rgb}{0,0.5,0}

\usepackage{booktabs} 
\hypersetup{colorlinks, linkcolor={blue}, citecolor={blue}, urlcolor={blue}} 
\usepackage{xspace}

\title[Short GRB merger ejecta emission search using GMRT]{Search for merger ejecta emission from late-time radio observations of short GRBs using GMRT}

\author[Ankur Ghosh et al.]{Ankur Ghosh$^{1,2}$\thanks{E-mail: ghosh.ankur1994@gmail.com (AG)},
C. S. Vaishnava$^3$,
L. Resmi$^3$,
Kuntal Misra$^1$,
K. G. Arun$^4$,
\newauthor
Amitesh Omar$^{1,5}$,
N. K. Chakradhari$^2$
\\
$^{1}$Aryabhatta Research Institute of observational sciencES (ARIES), Manora Peak, Nainital 263 001, Uttarakhand, India\\
$^{2}$School of Studies in Physics and Astrophysics, Pandit Ravishankar Shukla University, Raipur 492010, Chattisgarh, India\\
$^{3}$Indian Institute of Space Science and Technology, Trivandrum 695547, Kerala, India\\
$^{4}$Chennai Mathematical Institute, Siruseri, Chennai 603103, Tamilnadu, India\\
$^{5}$SPASE, Indian Institute of Technology, Kanpur 208016, Uttar Pradesh, India
}

\date{Accepted XXX. Received YYY; in original form ZZZ}

\pubyear{2015}

\begin{document}
\label{firstpage}
\pagerange{\pageref{firstpage}--\pageref{lastpage}}
\maketitle

\begin{abstract}
In some cases, the merger of two neutron stars can produce a rapidly rotating and highly magnetised millisecond magnetar. A significant proportion of the rotational energy deposited to the emerging ejecta can produce a late-time radio brightening from interacting with the ambient medium. Detection of this late-time radio emission from short GRBs can have profound implications for understanding the physics of the progenitor. We report the radio observations of five short GRBs - 050709, 061210, 100625A, 140903A, and 160821B using the legacy Giant Metrewave Radio Telescope (GMRT) at 1250, 610, and 325 MHz frequencies and the upgraded-GMRT (uGMRT) at band 5 (1050-1450 MHz) and band 4 (550-900 MHz) after $\sim$ $2 - 11$ years from the time of the burst. The GMRT observations at low frequencies are particularly important to detect the signature of merger ejecta emission at the peak. These observations are the most delayed searches associated with some GRBs for any late-time low-frequency emission. We find no evidence for such an emission. We find that none of these GRBs is consistent with maximally rotating magnetar with a rotational energy of $\sim 10^{53}\, {\rm erg}$. However, magnetars with lower rotational energies cannot be completely ruled out. Despite the non-detection, our study underscores the power of radio observations in the search for magnetar signatures associated with short GRBs. However, only future radio observatories may be able to detect these signatures or put more stringent constraints on the model.
\end{abstract}




\section{Introduction}
\label{introduction}
Double neutron star (DNS) or neutron star--black hole mergers have been argued to be the most promising progenitors for short-duration Gamma-Ray Bursts (short GRBs; \citealt{1986ApJ...308L..43P,1992ApJ...395L..83N,1999A&A...344..573R}). The Gravitational Waves (GWs) discovery from the DNS merger GW~170817 and simultaneous observation of a short GRB~170817A, along with the discovery of its electromagnetic counterparts in various bands, have revolutionised the era of multi-messenger astronomy \citep{2017ApJ...848L..13A,2017ApJ...846L...5G,2017ApJ...848L..15S} and has strengthened the hypothesis that short GRBs result from the merger of compact objects.

However, there are uncertainties about the final phase of the merger as GW observations are not sensitive to the post-merger dynamics of neutron star mergers (see, for example, Fig. 1 of \citealt{2013CQGra..30l3001B}) given the current sensitivities of the detectors. Numerical simulations have shown that the merger remnant may form a rapidly spinning supra-massive, highly magnetised neutron star (magnetar) before collapsing to a black hole \citep{2011AAS...21830201O, 2013ApJ...771L..26G}. Whether the situation occurs or not depends crucially on the resultant mass of the remnant and the highly uncertain equation of states (EoS) of dense neutron stars \citep{2011AAS...21830201O, 2016MNRAS.458.1660L, 2016ApJ...820...28O, 2020ApJ...890...99L}. With the recent discovery of millisecond pulsar MSP J0740+6620 having the mass of $2.14^{+0.10}_{-0.09}$ $M_{\odot}$ \citep{2020NatAs...4...72C}, this mass is often used as the lower limit of maximum neutron star mass. For a binary mass < 3 $M_{\odot}$, a long-lived supra-massive neutron star remnant could be formed \citep{2006Sci...311.1127D, 2013ApJ...771L..26G} before collapsing to a black hole. 

In the case of a DNS merger, the resultant product will be rapidly rotating with a spin period close to the centrifugal breakup value (P $\sim$ 1 ms). The remnant could also acquire a strong magnetic field $\geq 10^{14} - 10^{15}$ G, which may be enhanced by the Kelvin-Helmholtz instabilities and the dynamo activity \citep{1992ApJ...392L...9D, 1992Natur.357..472U, 2006Sci...312..719P, 2015PhRvD..92l4034K, 2017MNRAS.471.1879G}. During the merger, when the neutron stars are tidally disrupted, mass is thrown out with sub-relativistic velocities, which is expected to undergo {\it r}-process nucleosynthesis and produce {\it UV/opt/IR} emission resulting in a `kilonova' \citep{1998ApJ...507L..59L, 2017ApJ...848L..27T, 2019LRR....23....1M}. Signatures of kilonova were seen in the optical - IR afterglow light curves of a few short bursts, including GRB~170817A/AT2017gfo \citep{2017PASA...34...69A, 2017Sci...358.1565E, 2017Natur.551...75S, 2017ApJ...848L..27T, 2017ApJ...848L..24V, 2018ApJ...861L..12L}. Other short GRBs identified with kilonova signatures are GRB~050709 \citep{2016NatCo...712898J}, GRB~070809 \citep{2020NatAs...4...77J}, GRB~130603B \citep{2013ApJ...774L..23B, 2013Natur.500..547T}, and GRB~160821B \citep{2016GCN.20222....1T, 2019ApJ...883...48L}. The ejecta responsible for the kilonova eventually expands into the ambient medium, driving a shock and producing radio synchrotron emission \citep{2011Natur.478...82N, 2013MNRAS.430.2121P, 2015MNRAS.450.1430H}. This emission is also called `kilonova afterglow' in the literature \citep{2019MNRAS.487.3914K}.

However, the nature of the merger remnant will also play a crucial role in the evolution of the shock. In the case of a black-hole central engine, the velocity of the shock is limited by the initial kinetic energy of the kilonova ejecta. The flux will be highly boosted if the resultant product of the merger is a rapidly rotating and highly magnetised supra-massive neutron star or a magnetar \citep{2014MNRAS.437.1821M} as opposed to a black hole formed by direct collapse. This type of neutron star remnant is very unstable, and it generally tends to convert its rotational energy into kinetic energy via spin-down process \citep{2001ApJ...552L..35Z, 2008MNRAS.385.1455M, 2014ApJ...785L...6S}. This energy is directly imparted to re-energise the kilonova ejecta and accelerate it to mildly relativistic velocities. The interaction of the re-energised ejecta with the ambient medium produces a brighter radio emission visible from cosmological distances. The radio emission usually peaks at $\sim 5-10$ years since burst at $\sim 600$~MHz for typical values of the magnetar rotational energy, ejecta mass, and ambient medium density. Under favourable conditions, the radio emission can be visible in the timescale of years at lower frequency frequencies \citep{2014MNRAS.437.1821M}.

Prior to our work, there were other studies \citep{2014MNRAS.437.1821M, 2016ApJ...819L..22H,2016ApJ...831..141F, 2019ApJ...887..206K, 2020ApJ...902...82S, 2021MNRAS.500.1708R} to search for late-time merger ejecta emission using the VLA and the ATCA observations. \citet{2014MNRAS.437.1821M} started the search with seven GRBs using VLA at 1.4 GHz within the rest frame period of $\sim 0.5 - 2$ years since the burst trigger time, which resulted in upper limits of $\sim 200 - 500$ $\mu$Jy. They constrain the number density values to be $n_0 \leq 10^{-1} {\rm cm^{-3}}$. Another study by \citet{2016ApJ...819L..22H} targeted observations of two kilonova-associated short GRBs (GRB 060614; \citealt{2015ApJ...811L..22J, 2015NatCo...6.7323Y} and GRB 130603B; \citealt{2013ApJ...774L..23B, 2013Natur.500..547T, 2013ApJ...775L..19J, 2014arXiv1401.2166P}) with the ATCA 2.1 GHz and the VLA 6 GHz. These observations resulted in an upper limit of 150 and 60 $\mu$Jy observed at $7.9$ and $1.3$ years since the burst, respectively. Later, \citet{2016ApJ...831..141F} observed a sample of nine bursts with the VLA at 6 GHz and reported upper limits of $18 - 32$ $\mu$Jy. Their study constrained the number density value to be $n_0 \leq 10^{-3} {\rm cm^{-3}}$ for a large value of rotational energy $E_{\rm rot} \sim 10^{53}$ erg and ejecta mass of $M_{\rm ej} \sim 10^{-2} M_{\odot}$. The sample in \citet{2020ApJ...902...82S} consisted of nine GRBs (at z $\leq$ 0.5) observed using VLA at 6 GHz within $2-13$ years since the burst. This study quoted upper limits of $3 - 19.5$ $\mu$Jy on the flux density. In \citet{2021MNRAS.500.1708R}, a large sample of short GRBs was observed with ATCA at 2.1 GHz and VLA at 3 and 6 GHz. This work rules out the presence of a powerful magnetar as a merger remnant. \citet{2020ApJ...902...82S} and \citet{2021MNRAS.500.1708R} put stringent constraints on the number density values between 0.002 to 0.2 cm$^{-3}$ for rotational energy $\leq 10^{52}$ erg and ejecta mass $\leq$ 0.12 $M_{\odot}$.

Our study presents radio observations of five short GRBs at the rest-frame timescale of $\sim 2 - 11$ years after the burst to search for merger ejecta emission at late times. As the spectrum of the late-time merger ejecta emission is expected to peak in the MHz frequency range (around 600 MHz), we performed an extensive search using the legacy GMRT and uGMRT at frequencies below 1.4 GHz. To model the light curve, we employ a model of energy injection by a magnetar central engine, following \citet{2014MNRAS.437.1821M} and account for Doppler correction and transition into a deep Newtonian phase following \cite{2003MNRAS.341..263H}. Further, our model also accounts for the evolution of the bulk Lorentz factor. Although we do not find any evidence of radio emission in the short GRBs in our sample, we derive crucial limits on the energetics of the bursts and the parameters of the ambient medium.

In section \ref{sec:sample_selection}, we discuss the selection criteria of the short GRBs in our sample, followed by the radio data acquisition and reduction procedure in Section \ref{reduction}. Section \ref{sec:magnetar} introduces the magnetar model and the changes incorporated in our model compared to the previous studies. The results of this work and the summary are presented in Sections \ref{results} and \ref{summary}, respectively.

\section{Sample Selection}
\label{sec:sample_selection}
\begin{table*}
\centering
\caption{Details of the GRBs in our sample}
\setlength{\tabcolsep}{14pt}
\smallskip
\begin{tabular}{l c c c c c c}
\hline \hline
GRB Name  & Redshift       & $T_{90}$      & Fluence  	& $E_{\rm iso}$     	& Counterpart  &  References    \\
          &                &              &          	&              	& detected &  \\
     & (z)           & (sec) 	      &     (erg $cm^{-2}$)        &    (erg)                 &   &   \\
 \hline    
050709  &   0.1606 &  $0.07\pm0.01$    & ($7.1\pm1.5)\times10^{-7}$ &  $2.7_{-0.3}^{+1.1} \times 10^{49}$   & X, O & 1 , 2, 3  \\

061210  &   0.4095   & $0.047$         & ($1.1\pm0.2)\times10^{-6}$ &  $4.6\times 10^{50}$ & X & 4, 5 \\

100625A &  0.4520    & $0.33\pm0.03$   & ($2.2\pm0.3)\times10^{-7}$ & $\sim 10^{51}$        & X   &  6, 7, 8  \\

140903A &  0.3510     &  $0.30\pm0.03$   &  ($1.4\pm0.1)\times10^{-7}$ & ($5.9\pm0.3)\times10^{49}$              & X, O, R & 9, 10, 11, 12 \\

160821B & 0.1613   &  $0.48\pm0.07$    &  ($2.5\pm0.9)\times10^{-6}$ &  ($2.1\pm0.2)\times10^{50}$  & X, O, R  &   13, 14, 15\\
\hline
\end{tabular}
\footnotesize{Note: X = X-ray, O = optical, R = radio. $1$ - \citet{2005AAS...207.1904V}, $2$ - \citet{2005GCN..3570....1B}, $3$ - \citet{2005GCN..3570....1B}, 4 - \citet{2006GCN..5912....1C}, 5 - \citet{2006GCN..5917....1U}, 6 - \citet{2021ApJ...912...14Y}, 7 - \citet{2013MNRAS.430.1061R}, 8 - \citet{2013ApJ...769...56F}, 9 - \citet{2017ApJ...843L...1L}, 10 - \citet{2019ApJ...872..114S}, 11 - \citet{2014GCN.16774....1C}, 12 - \citet{2014GCN.16768....1P}, 13 - \citet{2017ApJ...835..181L}, 14 - \citet{2016GCN.19844....1P}, 15 - \citet{2016GCN.20222....1T}}
\label{tab:sample}
\end{table*}

In previous studies by \citet{2014MNRAS.437.1821M,2016ApJ...819L..22H, 2016ApJ...831..141F, 2020ApJ...902...82S} and \citet{2021MNRAS.500.1708R}, the fields of short GRBs were observed a few years after the burst in 1.4, 3, and 6 GHz with the VLA and 2.1 GHz with the ATCA. In our study, we implemented changes in the observing strategy. Our objective was to use the legacy GMRT in 325 and 610 MHz and the uGMRT in band 5 and band 4 for this study as the spectrum of the late-time merger ejecta emission is expected to peak in the lower frequency radio regime. The uGMRT with much wider bandwidth (while the legacy GMRT worked with the bandwidth of 32 MHz, the bandwidth was improved up to 400 MHz in uGMRT) and improved sensitivity than the legacy GMRT, helps to reduce the rms of the image.

The sample selection and observational criteria are mentioned below.

1. We chose nearby bursts (z $<$ 0.5) around 2 to 11 years since the burst for the proposed period of GMRT observations. 18 bursts fulfilled this criterion.

2. We extrapolated the VLA limits of the GRBs available in \citet{2016ApJ...819L..22H, 2016ApJ...831..141F} to the GMRT bands and found that five GRBs (080905A, 050724, 130603B, 090515, and 150120A) were below the GMRT detection limit and were thus removed from the sample.

3. We excluded bursts that did not show an X-ray plateau or, early extended emission (EE), or kilonova signature (reported at the time of sample selection). This resulted in eliminating GRBs~050509B, 060502A, 100206, and 070724A.

4. We excluded bursts for which afterglow modelling indicated the possibility of a low ambient density ($10^{-4}$ $cm^{-3}$) (GRBs~150101B and 061006).

5. The bursts (GRBs~061201 and 071227) beyond the declination limit ($-52^{\circ}$ to $+90^{\circ}$) of GMRT were excluded from our sample because of their inaccessibility with the GMRT. 

6. We also removed the ambiguously classified burst GRB~060614.

Afterwards, we included GRB~140903A, which had no reported X-ray plateau/EE/kilonova because it had a tight $n_0$ limit from afterglow observations. Thus, we ended up with five bursts (GRBs~050709, 061210, 100625A, 140903A, and 160821B) which were observed with the GMRT/uGMRT under our approved proposals (31\_109, PI: L. Resmi and 34\_096, PI: L. Resmi) in cycles 31 and 34. The properties of the individual bursts are summarised in Table \ref{tab:sample}.

Although at different frequencies and time scales, other telescopes observed some GRBs in this sample earlier. These observational details are listed in Table \ref{tab:other_observations}.

\begin{table*}
\caption{Other observations of the GRBs in this sample available in the literature}
\centering
\smallskip
\begin{tabular}{l c c c c c c}
\hline \hline
GRB        &   Telescope Name    	&  Frequency &  $T_{\rm obs}^{\dagger}$  &  Flux density$^{\ddagger}$ &  $L_{\nu}$ &   References  \\
                  &              	&  (GHz)  & (years)	      & ($\mu$Jy)   & (erg/s/Hz)  &  \\
\hline         
050709	& VLA  &	1.4	& 2.53     &  $<$ 349 & $ < 2.92 \times 10^{29}$  &  \citet{2014MNRAS.437.1821M}	\\
            & VLA   &	3.0 & 12.57   & $<$  31 & $< 2.60 \times 10^{28}$ & \citet{2016NatCo...712898J} 	\\

061210        & VLA  & 6.0	& 7.85   & $<$ 27 & $< 2.33 \times 10^{29}$ &  \citet{2019ApJ...887..206K}	\\
140903A  &  ATCA &  2.1 & 1.86  & $<$ 153  & $<8.80 \times 10^{29}$ & \citet{2016ApJ...827..102T}  \\
            & VLA &  6.0 &  5.99 &  $<$ 17.5 & $< 1.01 \times 10^{29}$ &\citet{2016ApJ...827..102T} 	\\
160821B  & VLA & 6.0  & 2.77   & $<$ 6 & $< 5.08 \times 10^{27}$&\citet{2019MNRAS.489.2104T} \\
\hline

\multicolumn{6}{l}{$^{\dagger}$ time is calculated since burst trigger time.}\\
\multicolumn{6}{l}{$^{\ddagger}$ Upper limits correspond to 3-$\sigma$  confidence.}\\

\end{tabular}
\label{tab:other_observations}      
\end{table*}

\section{Observations and Data Reduction}
\label{reduction}

\begin{figure}
\includegraphics[width=\columnwidth]{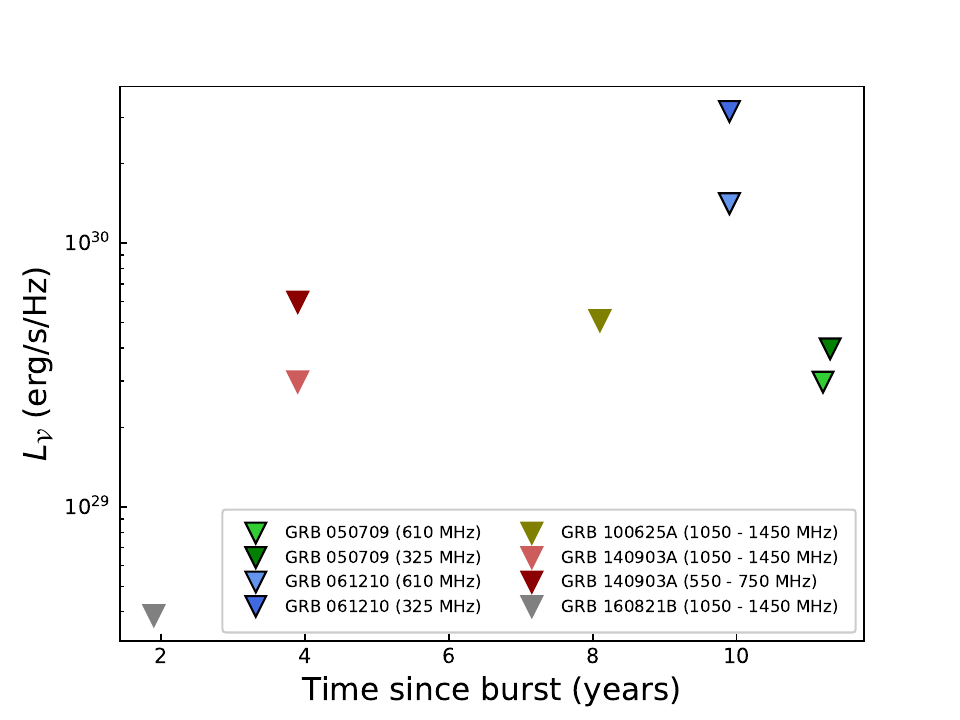}
\caption{Spectral luminosity of legacy and uGMRT observations for the five short GRBs in our sample. The inverted triangles with a black border represent the legacy GMRT 3-$\sigma$ upper limits, and those without a border represent uGMRT 3-$\sigma$ upper limits.} 
\label{upper_limits}
\end{figure}

The selected sample of five GRBs (section \ref{sec:sample_selection}) was observed in radio frequencies using the legacy GMRT \citep{1991ASPC...19..376S, 2005ICRC...10..125A} and uGMRT \citep{2017CSci..113..707G}. Two GRBs (GRBs 050709 and 061210) were observed with the legacy GMRT in 610 and 325 MHz (31\textunderscore109, PI: L. Resmi). Further, we observed three GRBs (GRBs 100625A, 140903A, and 160821B) with the uGMRT (34\textunderscore096, PI: L. Resmi), either in band 5 or band 4 or both bands. The observational details of all five GRBs are given in Table \ref{tab:gmrt_observations}.

\begin{table*}
\caption{Details of the legacy-GMRT/uGMRT observations}
\centering
\smallskip
\begin{tabular}{l c c c  c c c}
\hline \hline
GRB         &Frequency    &   UT Date     	& $T_{\rm obs}^{\dagger}$      &    Flux density$^{\ddagger}$  & $L_{\nu}$ & \\
            & (MHz)      &              	&  (years)  	       (min)   & ($\mu$Jy) & (erg/s/Hz) & \\
\hline         
050709	& 610    &	2016-11-21 09 00 00	& 11.2        & $<$ 360  & $ < 2.97 \times 10^{29}$ &   legacy GMRT\\
            & 325    &	2017-01-21 04 30 00 & 11.3     & $<$ 481 & $< 3.97 \times 10^{29}$  &  	legacy GMRT\\

061210        & 610   & 2016-11-29 18 15 00	& 9.9   & $<$ 165  & $<1.41 \times 10^{30}$ & legacy GMRT 	\\
      & 325  & 2016-11-28 16 00 00	& 9.9     & $<$ 369 &< $3.15 \times 10^{30}$ & legacy GMRT	\\
100625A  & band 5 &  2018-07-24 18 45 00  & 8.1     & $<$ 45.6  & $< 5.08 \times 10^{29}$ & $uGMRT^{\star}$\\
140903A  & band 5  &  2018-07-24 12:30:00 & 3.9   & $<$ 52.5 & $< 2.98 \times 10^{29}$ & uGMRT\\
            &      band 4 &   2018-07-24 08 30 00	& 3.9   &  $<$ 105 & $<5.97 \times 10^{29}$ & uGMRT 	\\
160821B  & band 5 & 2018-07-24 15 30 00  & 1.9    & $<$ 46.5 &< $3.87 \times 10^{28}$ & uGMRT\\
\hline
\multicolumn{6}{l}{$^{\dagger}$ time is calculated since burst trigger time.}\\
\multicolumn{6}{l}{$^{\ddagger}$ Upper limits correspond to 3-$\sigma$  confidence.}\\
\multicolumn{6}{l}{$^{\star}$ uGMRT frequencies are the average frequencies (band 5 - 1250 MHz and band 4 - 650 MHz) at which the maps were created.}\\
\end{tabular}
\label{tab:gmrt_observations} \end{table*}

The legacy GMRT data was processed using the standard flagging, calibration, and imaging procedure in the Astronomical Image Processing System (AIPS; \citealt{2003ASSL..285..109G}). The uGMRT data was processed in Common Astronomy Software Application (CASA) v. 5.5 and a customised pipeline developed in CASA by Ishwar-Chandra et al. (2023, in preparation). In legacy GMRT and uGMRT, flux and phase calibrators were observed along with the source for calibrating the data. After detailed flagging and standard calibration, the data was visually inspected to identify bad data, which was discarded. Cleaning and imaging were done using a cell size of 1.0" and 1.5" for 610 and 325 MHz of legacy GMRT, respectively. Similarly, we used cell sizes of 0.5" and 1.0" for band 5 and band 4 of uGMRT. Five times phase-only self-calibration and three times amplitude \& phase self-calibration have been done to improve the image quality and remove the artefacts. 

At the location of the X-ray afterglow of these GRBs, we do not detect a radio source in our observations. We used the AIPS/IMSTAT task in each image plane on a source-free region close to the region of interest (in this case, the GRB X-ray afterglow location) to estimate the map rms. The values of spectral luminosity are given in Table \ref{tab:gmrt_observations} and shown in figure \ref{upper_limits}.

\section{Late-time radio emission from magnetar powered merger ejecta}
\label{sec:magnetar}

We now summarise our model for the merger ejecta from magnetars and corresponding late-time radio emission. Simulations have shown that the dynamical mass ejection occurs in the case of DNS mergers while a comparable or larger amount of mass may be ejected in the outflows from the accretion disk \citep{2013PhRvD..87b4001H}. The modeling of kilonova emission from GRB~130603B \citep{2013ApJ...774L..23B, 2013Natur.500..547T} led to the inference that a higher amount of mass ($\sim 0.03 - 0.08 M_{\odot}$) can be associated with such dynamical ejection. We next examine the dynamics of this ejecta and how it gives rise to radio emission.

\subsection{Shock dynamics}
\label{secdyn}

We consider a simple ejecta model with initial velocity $\beta_0$ expanding into a homogeneous ambient medium of density $n_0$. Due to energy injection from the magnetar, the merger ejecta can achieve sub-relativistic velocities. Therefore, adopting an exhaustive dynamical model applicable to asymptotic ultra-relativistic and non-relativistic regimes is crucial. This is more important because the peak flux is expected at the deceleration of the fireball. We follow the comprehensive treatment developed by \cite{PeerDynamcis2012} for the evolution of the blast wave as the ejecta interacts with the surrounding medium. 

Following \cite{PeerDynamcis2012}, we solve the equation for the evolution of bulk Lorentz factor ($\Gamma$) with the swept-up mass $m$

\begin{equation}
\label{gamma_evo}
    \frac{d\Gamma}{dm} = -\frac{\hat{\gamma}(\Gamma^2 - 1) -(\hat{\gamma} - 1) \Gamma \beta^2}{M_{\rm ej} + m[2 \hat{\gamma}\Gamma - (\hat{\gamma} - 1) (1 + \Gamma^{-2})]},
\end{equation}
where $\beta$ is the velocity of the ejecta divided by the speed of light c, $\gamma(\beta)$ is the Lorentz factor of the blast wave, $\hat{\gamma}$ is the adiabatic index of the shocked gas, and $M_{\rm ej}$ is the mass of the dynamic ejecta from the merger. We used the same functional form for $\hat{\gamma} (\gamma \beta)$ given in \cite{PeerDynamcis2012}. The mass swept up from the ambient medium, $m$, is given by.
\begin{equation}
 \label{dmeqn}
dm = 4 \pi r^2 n_0 m_p,
\end{equation}
where $r$ is the radius of the blast-wave, $n_0$ is the ambient density, and $m_p$ is the mass of proton. 
Blast-wave radius and the observed time $t_{\rm obs}$ are related by,
\begin{equation}
\label{timeeqn}
\frac{d}{dr} \frac{t_{\rm obs}}{1+z}= \frac{(1-\beta)}{\beta c},
\end{equation}
where $z$ is the redshift of the GRBs and $c$ is the velocity of light.

We assume that the entire magnetar rotational energy $E_{\rm rot}$ is converted to the kinetic energy of the merger ejecta of mass $M_{\rm ej}$, i.e., $E_{\rm rot} = E_{\rm kin} = (\gamma_0 -1) M_{\rm ej} c^2$, where $\gamma_0$ is the Lorentz factor corresponding to the initial velocity. Rotational energy of the magnetar depends on its period $P$ as $E_{\rm rot} = 3 \times 10^{52} {\rm erg} \left( {\frac{P}{1 \rm{ms}} } \right)^{-2}$, for a mass of $1.4 M_{\odot}$. Therefore, for a magnetar of $\sim 1 \rm ms$ rotation period, the ejecta can achieve mildly relativistic velocities given by 
\begin{equation}
 \gamma_0-1  = 1.7 \left( \frac{E_{\rm rot} }{ {3 \times 10^{52} \rm{erg}} } \right) \left( \frac{ M_{\rm ej}}{0.01M_\odot} \right)^{-1}. 
\end{equation}

Using this assumption, we solve for the above three simultaneous differential equations and obtain the temporal evolution of the radius $r(t)$ and velocity $\beta(t)$ of the blast wave. 
Since we aim to obtain limits on the energy input from a potential magnetar central engine, our basic parameters for the dynamics are $E_{\rm rot}$ and $M_{\rm ej}$ in addition to $n_0$.

We do not consider the initial phase where the dynamical ejecta is accelerated by the magnetar inclusion of this phase can modify the flux before and around the deceleration epoch (see \citet{2020ApJ...890..102L} for details). In addition, our model does not include the impact the jet preceding the ejecta has created in the ambient medium, leading to radio flares well before the epoch of deceleration of the ejecta \citep{2020MNRAS.495.4981M}. Although both of these can influence the light curve, including them is beyond the scope of this paper.

\subsection{Synchrotron emission}
\label{secradn}
Afterglow radiation emerges from the region behind the shock front (shock downstream), where the bulk kinetic energy of the outflow is converted to thermal energy. So, the synchrotron flux downstream of the shock depends on the fraction of shock thermal energy in non-thermal electrons ($\epsilon_e$) and magnetic field ($\epsilon_B$). In addition, the flux depends to some degree on the power-law index $p$ characterising the distribution of electrons in energy space. Changes in the value of $p$ do not influence the result significantly, but that on $\epsilon_e$ and $\epsilon_B$ will. However, as the observations have only resulted in upper limits, it is crucial to bring down the free parameters of the problem. Therefore, considering near equipartition in the shock downstream, we assume $\epsilon_e \sim \epsilon_B \sim 0.1$ \citep{2017MNRAS.472.3161B}. We also assume that the electrons are distributed as a power-law of index $p=2.2$ in energy.

To obtain the synchrotron spectral energy distribution, we first calculate the downstream magnetic field as $B^2/8 \pi = 4\epsilon_Bn_0 \;  \gamma (\gamma-1)$ and the minimum Lorentz factor $\gamma_m$ of the electron distribution as $\gamma_m = 1+\epsilon_e (p-2)/(p-1) \; (m_p/m_e)\; (\gamma -1 )$ \citep{2006ApJ...653..454P}. 

At sufficiently late times, the minimum Lorentz factor approaches unity and remains a constant. To circumvent numerical artefacts appearing from assuming $\gamma_m=1$, we terminate the evolution of $\gamma_m$ at $2$. For example, for $\epsilon_e = 0.1$, and $p=2.2$, $\gamma_m$ drops to $2$ when $(\gamma(t)-1) = \beta(t)^2/2 = 0.02$. For $E_{\rm rot} = 3\times10^{52}$~erg, $M_{\rm ej}$ = 0.05 $M_\odot$, and $n_0=0.01$ $cm^{-3}$, this value of $\gamma(t)$ occurs at $\sim$ 106 years since the merger. 

We follow Rybicki \& Lightman \citep{1979rpa..book.....R} to calculate the characteristic synchrotron frequency $\nu_m$ corresponding to the electron of Lorentz factor $\gamma_m$, and the flux density $f_m$ at $\nu_m$, given by,

\begin{equation}
\label{gamma_m}
    \begin{split}
\nu_m& = \frac{4190~{\rm MHz}}{(1+z)} \;  \frac{B(t)}{\rm mG} \; \gamma_m(t)^2\; \gamma(t), \\
 f_m & = 240~{\rm \mu Jy} \; (1+z) \; n_{0,-2} \; \frac{B(t)}{\rm mG}\,  \gamma(t) \; \frac{r_{\rm pc}(t)^3}{d_{\rm L,Gpc}^2}.
    \end{split}
\end{equation}

where $r_{\rm pc}(t)$ is the radius in parsec and $d_{\rm L,Gpc}$ is the luminosity distance in gigaparsec.

Synchrotron self-absorption may be relevant for MHz frequencies for some parts of the parameter space. To estimate the self-absorption frequency ($\nu_a$), we calculate the rest-frame optical depth $\tau_{\nu_m}$ at frequency $\nu_m$ from \citet{1979rpa..book.....R}. We define $\nu_a$ as $\tau_{\nu=\nu_a} = 1$, where $\tau_{\nu} \propto \nu^{-5/3}$ for $\nu < \nu_m$ and $\propto \nu^{-(p+4)/2}$ for $\nu > \nu_m$. Resulting equations for $\nu_a$ for $p=2.2$ are given by,

for $\nu_a <\nu_m$
\begin{equation}
\nu_a  =\frac{130~{\rm MHz}}{(1+z)} \; n_{0,-2}^{4/5} \; r_{\rm pc}^{3/5} \frac{\epsilon_{B}^{1/5}}{\epsilon_{e}} \frac{\gamma(t)}{(\gamma(t)-1)^{4/5}} \\
\end{equation}

and for $\nu_a > \nu_m$

\begin{equation}
\nu_a = \frac{398~{\rm MHz}}{1+z}\; n_{0,-2}^{7/10} \; r_{\rm pc}^{3/10} (\gamma(t)-1)^{7/10} \gamma(t)^{7/20} \epsilon_{B}^{7/20} \epsilon_{e}^{2/5}
\end{equation}
where $\gamma(t)$ is the bulk Lorentz factor of the blast-wave respectively, which we obtain from solving the simultaneous differential equations eq-\ref{gamma_evo}, \ref{dmeqn} and \ref{timeeqn}. At later epochs where the blast wave is non-relativistic, $r$ evolves as $t^{2/5}$. In the expressions for $\nu_m$ and $\nu_a$, an additional redshift factor from $r(t)$ corresponds to the cosmological time dilation.

\begin{figure*}
\begin{flushleft}
\includegraphics[width=0.99\linewidth]{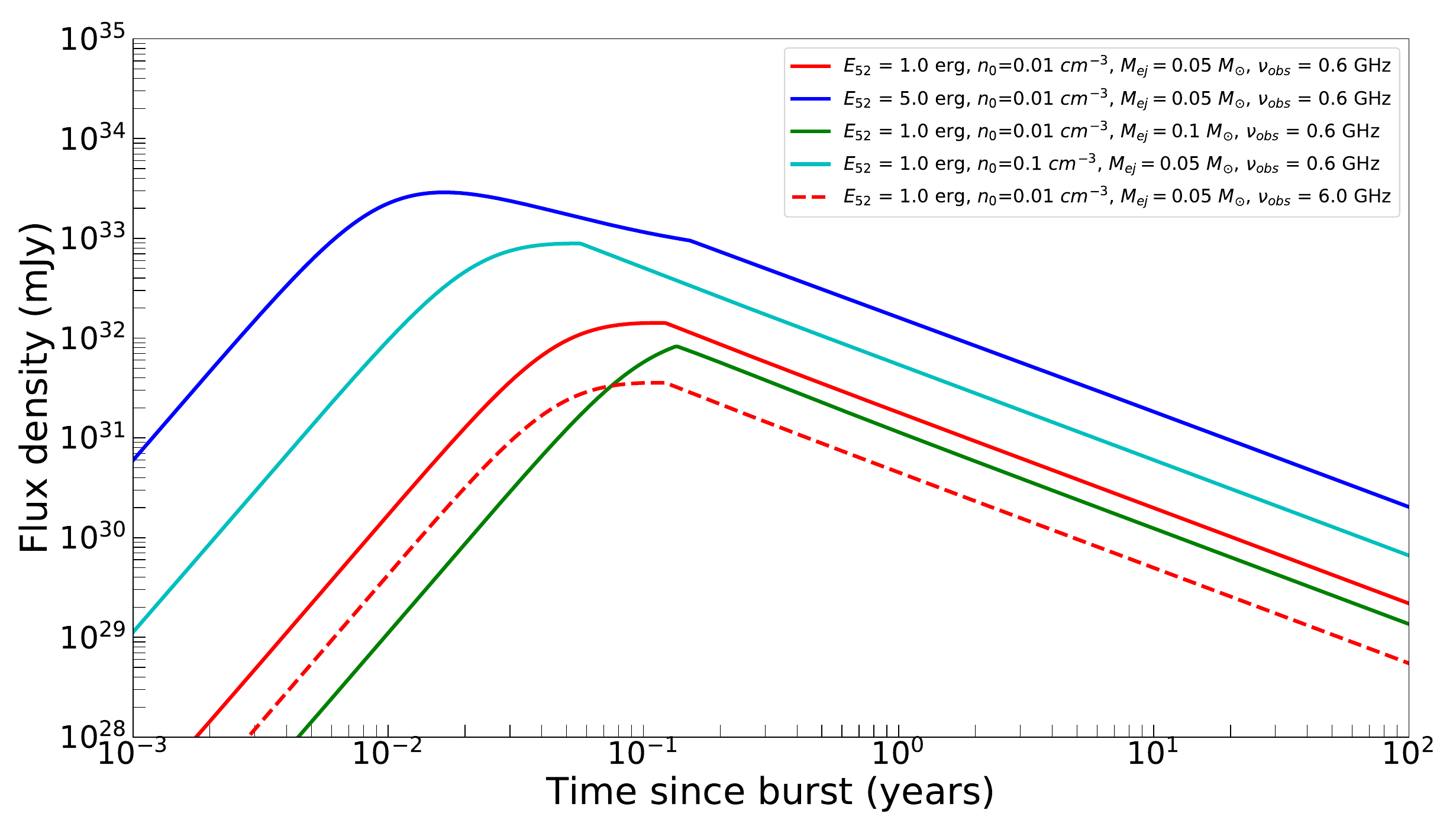}
\caption{Model light curves of at z = 0.1606 for a given Magnetar rotational energy ($E_{\rm rot}$), ejecta mass ($M_{\rm ej}$), and ambient medium density ($n_0$). We show the effect of variation of the different parameters on the light curve. As the fireball is optically thin, the low-frequency flux (solid red line) is higher than the high-frequency (red dash line).}
\label{multibandlc}
\end{flushleft}
\end{figure*}

The synchrotron spectrum then is calculated following \citet{2018pgrb.book.....Z}. In figure \ref{multibandlc}, we present light curves for various values of $E_{\rm rot}$, $M_{\rm ej}$, and $n_0$. The light curve peak corresponds to the fireball's deceleration or the transition from optically thick to the thin regime, whichever is later. All the light curves we have shown are optically thin throughout; hence, the peaks correspond to the epoch of deceleration. 

A higher $E_{\rm rot}$ or $n_0$ leads to higher downstream thermal energy and, therefore, a higher flux at a given epoch. An increase in any of the parameters reduces the deceleration time (i.e., the peak of the light curve). Larger energy leads to a higher initial velocity, reducing the time required for deceleration. Higher $n_0$ increases the rate at which mass is swept up, leading to a lower deceleration time. A larger $M_{\rm ej}$ leads to a lower initial velocity, resulting in a lower downstream thermal energy and, subsequently, lower flux. The deceleration epoch also increases because larger $M_{\rm ej}$ requires a higher mass to be collected for deceleration. The lower ejecta velocity further reduces the deceleration time.

\section{Results and discussion}
\label{results}

\begin{figure*}
	\includegraphics[width=0.99\linewidth]{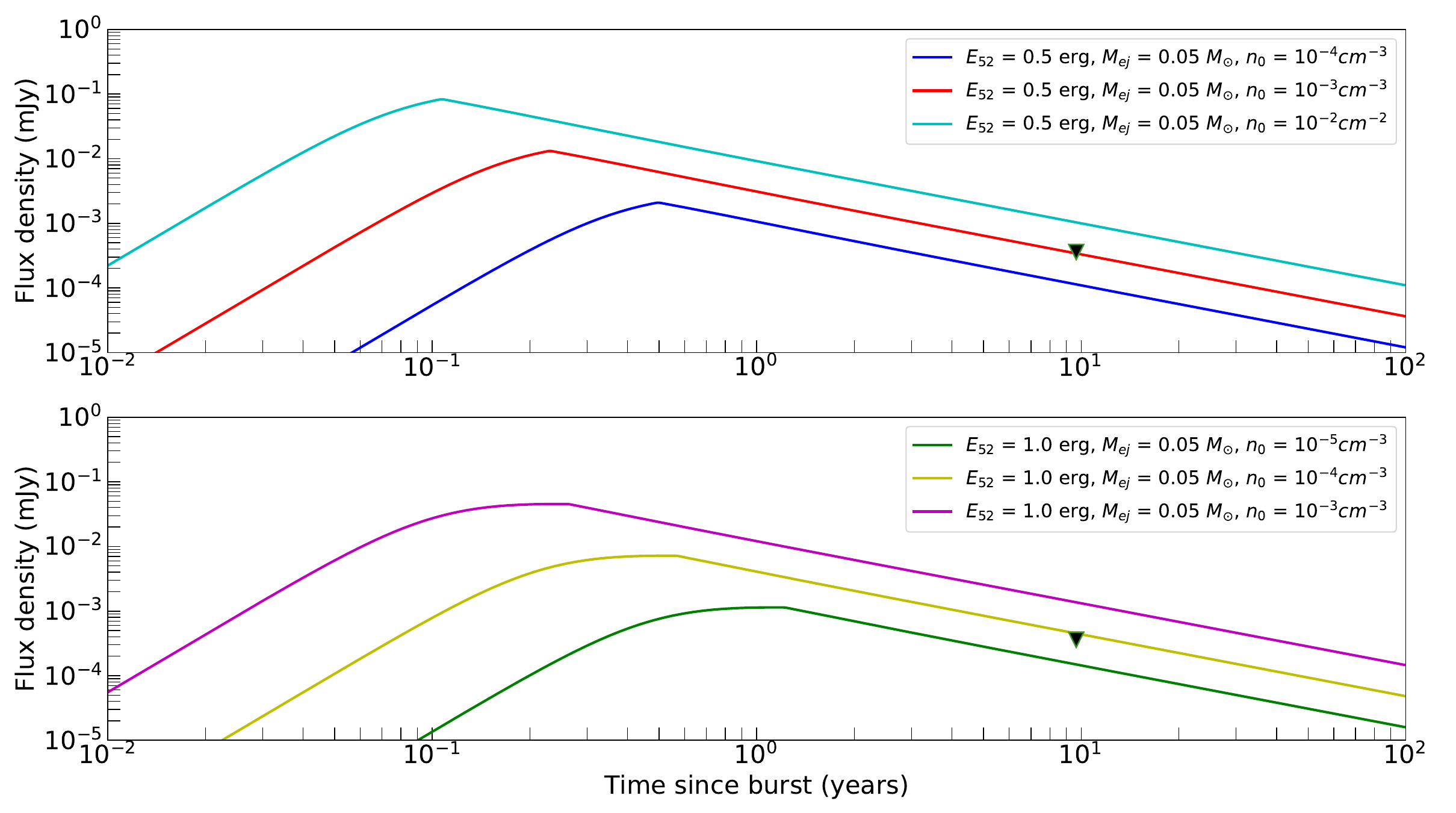}
    \caption{The upper and lower panel shows the model light curves of GRB 050709 (z = 0.1606) with the rotational energy of $5 \times 10^{51}$ erg and $10^{52}$ erg, respectively, in GMRT 610 MHz band. The black triangle denotes 3-$\sigma$ upper limits using GMRT.}
    \label{fig:GRB050709_lc}
\end{figure*}

Our data analysis did not lead to the detection of any radio emission from the GRBs we observed. Observed flux density limits for the merger remnant can be translated to an upper limit for the rotational energy of a potential magnetar remnant. In figure \ref{fig:GRB050709_lc}, we show the model light curves of GRB 050709 for band 4 and band 3 along with the GMRT 3$\sigma$ flux density limits of the GRBs as mentioned in table \ref{tab:gmrt_observations}. Along with the constraints on ambient density from afterglow data \citealt{2015ApJ...815..102F, 2016NatCo...712898J}, a long-lived magnetar with rotational energy above $5 \times 10^{52}$~erg can be ruled out for this burst. 

Next, we obtain the $E_{\rm rot}$--$n_0$ space for all bursts in our sample. This exercise becomes difficult due to a large number of model parameters and limited observational inputs. Therefore, as mentioned in section-\ref{secradn}, we fixed the microphysical parameters relevant to the radiation spectrum from the shock downstream at $\epsilon_e=0.1, \epsilon_B=0.1$, and $p=2.2$. 

Since the radio flux is directly proportional to both $\epsilon_B$ and $E_{\rm rot}$, a lower $\epsilon_B$ will lead to a shallower upper limit of $E_{\rm rot}$ for a given $n_0$ and $M_{\rm ej}$. We considered near equipartition between the magnetic field and electrons and fixed $\epsilon_B = 0.1$. A lower value of $\epsilon_B$ in the downstream is not unlikely and increases the allowed maximum value of $E_{\rm rot}$. On the other hand, if the ejecta mass reduces, $E_{\rm rot}$ upper limit gets tighter. 

For the ultra-relativistic shocks of GRB afterglows, a wide range of $\epsilon_B$ values are inferred \citep{2014ApJ...785...29S}. However, an equipartition magnetic field is commonly found in galactic SNRs \citep{2012A&ARv..20...49V}, where the shock is sub-relativistic. Theoretical calculations of magnetic field amplification due to shock-generated turbulence result in a higher value of $\epsilon_B$ \citep{1992ApJ...396..606K}. The late-time merger ejecta shocks are expected to be mildly relativistic or non-relativistic. Therefore, though there's a deviation in GRB afterglows, we chose to fix the $\epsilon_B$ at 0.1.

Numerical simulations provide a range of values for the dynamical ejecta mass \citep{2019ARNPS..69...41S}. However, we fix the ejecta mass at $0.05 M_{\odot}$, estimated for GW170817 by various authors \citep{2017ApJ...848L..17C, 2017Sci...358.1570D, 2018MNRAS.481.3423W}. This is a conservative upper limit to the dynamical ejecta mass from \cite{2019ARNPS..69...41S}.

For each burst, we calculate the model flux density $f_{\rm \nu}(t)$ on a grid of $E_{\rm rot}$ and $n_0$ as described below, compare it with the GMRT upper limits, and obtain the parameter space allowed by the observations. 

\begin{enumerate}
    \item {\it Rotational energy:} For the magnetar rotational energy, we choose a wide range from $5 \times 10^{51}$ erg to $10^{53}$ erg that is theoretically allowed as a function of the mass and the equation of state of the neutron star. The maximum rotational energy of a stable magnetar is around $10^{53}$ erg, after which the magnetar collapses into a black hole.
    \item {\it Number density:} As short GRBs occur in diverse environments, we consider a wide range for the number density, from $10^{-5}$ to $1$ ${\rm cm^{-3}}$. Whenever available, we have also considered the limits of $n_0$ from afterglow modeling (\citealt{ 2015ApJ...815..102F, 2016NatCo...712898J, 2016ApJ...827..102T, 2019MNRAS.489.2104T}).
\end{enumerate}

It must be noted that assumptions on $\epsilon_B$ and $M_{\rm ej}$ can significantly affect the results. For the optically thin spectral regime of $\nu_m \leq \nu \leq \nu_c$, we can arrive at the equation \ref{flux_ul} for the fireball in the non-relativistic phase.

\begin{equation}
\label{flux_ul}
    E_{\rm rot} \geq {\rm const} \frac{f_{\rm UL}^{2.6}}{\epsilon_e^{2.6}} \; \epsilon_B^{-2.0} \, M_{\rm ej}^{3.4}\, n_0^{-1.2},
\end{equation}

for $p=2.2$. $f_{\rm UL}$ is the 3-$\sigma$ flux density upper limit. The preceding constant factor is a combination of fundamental constants, frequency of observation, and distance to the burst.

\subsection{Allowed maximum rotational energy of magnetars}
\label{maxrot}

\begin{figure*}
	\includegraphics[width=0.85\columnwidth]{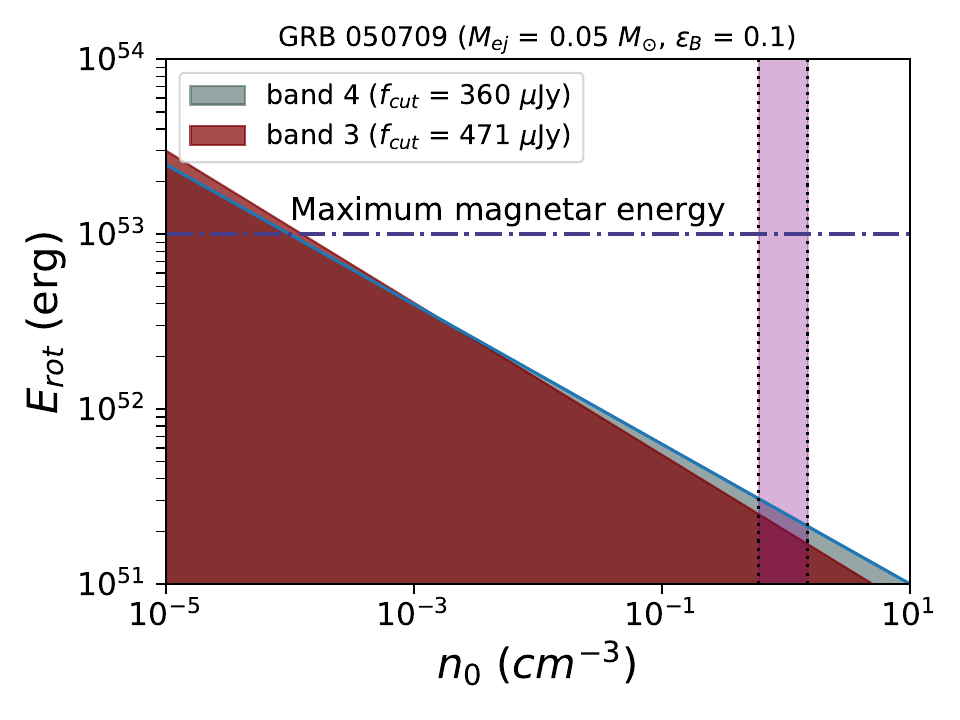}
	\includegraphics[width=0.85\columnwidth]{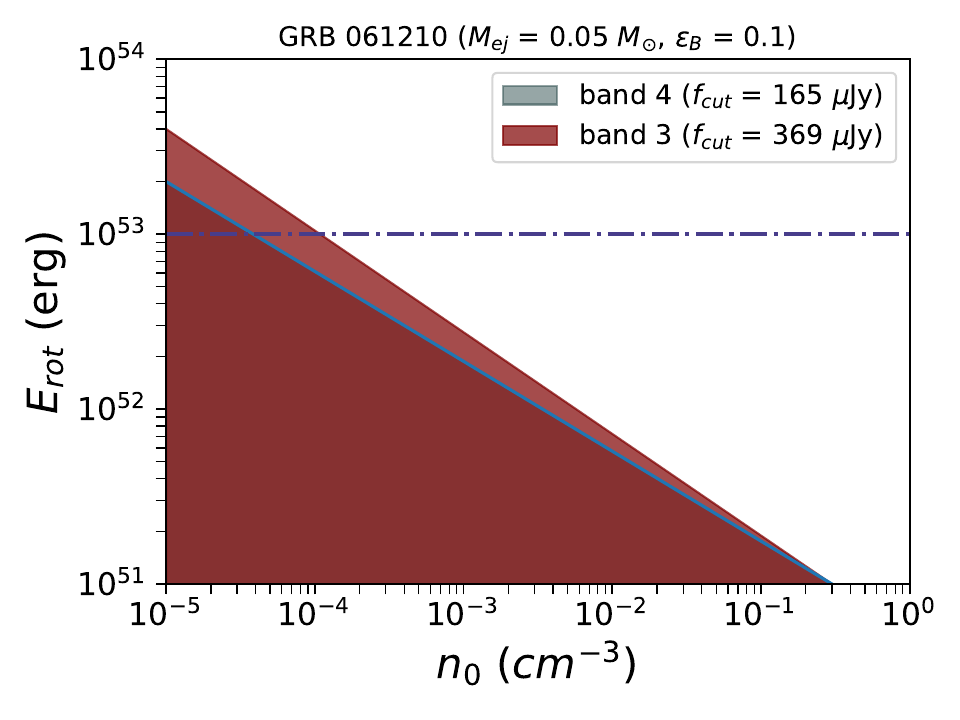}
	\includegraphics[width=0.85\columnwidth]{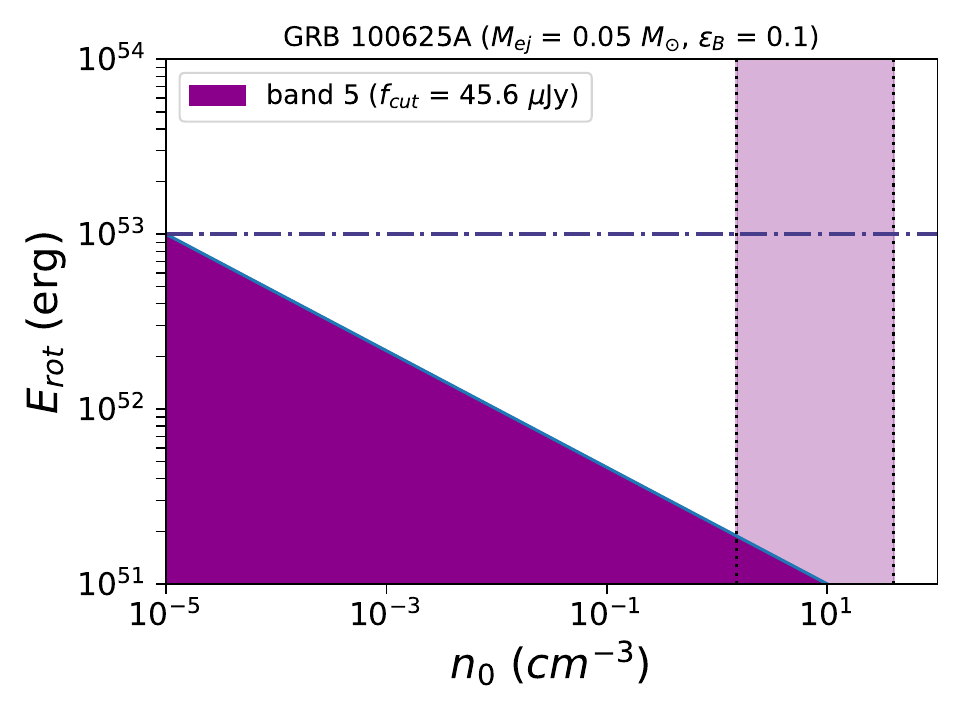}
	\includegraphics[width=0.85\columnwidth]{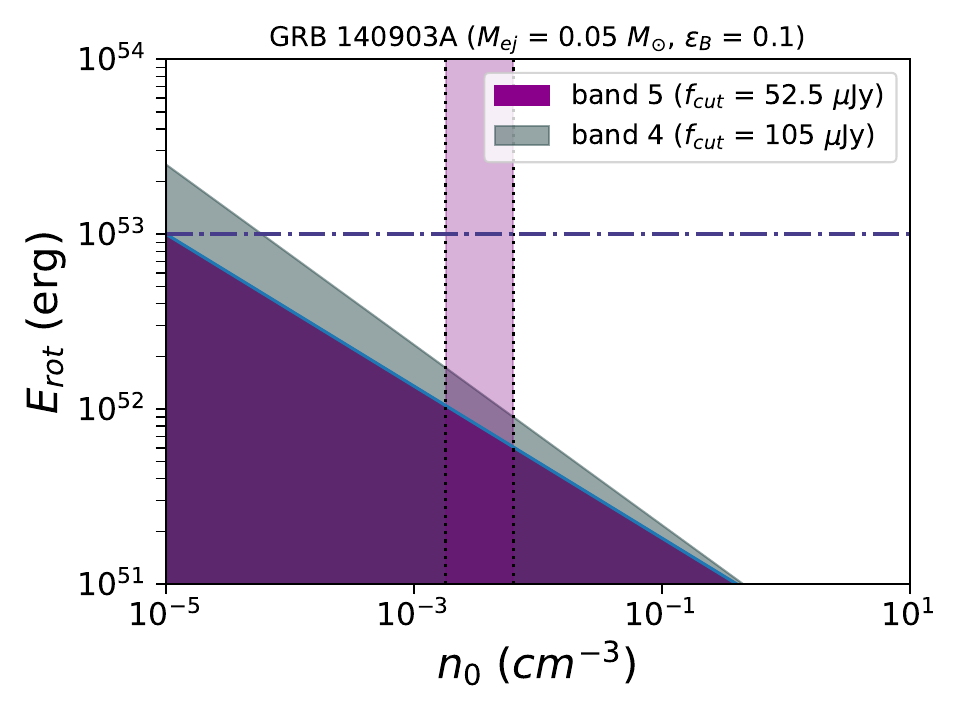}
	\includegraphics[width=0.85\columnwidth]{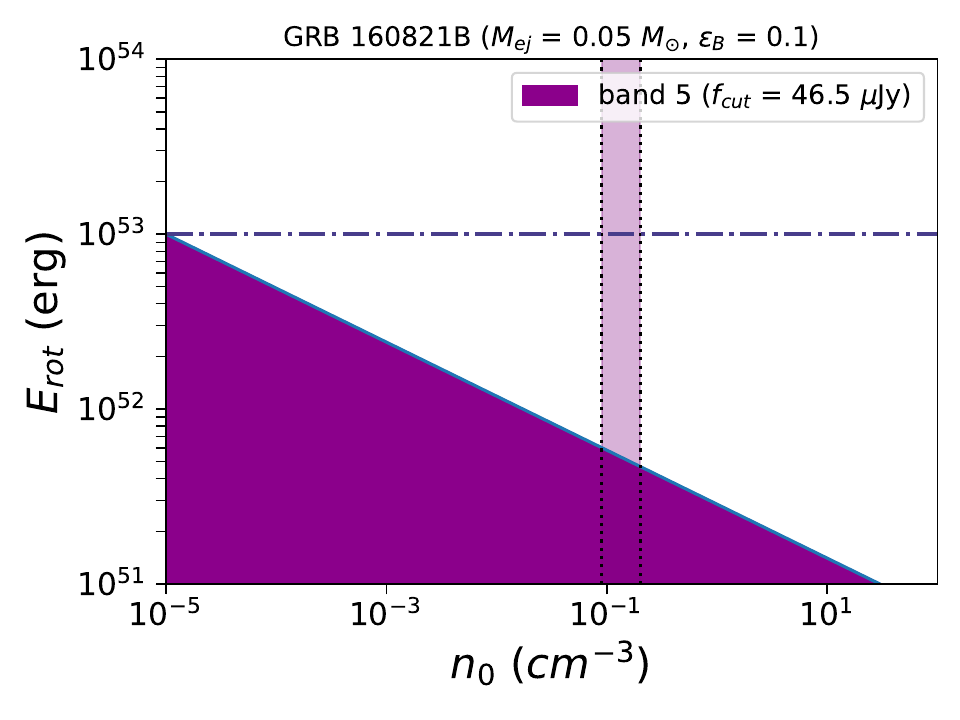}
    \caption{Rotational energy vs. number density parameter space for five GRBs observed with uGMRT band 5, band 4, and legacy GMRT 610 MHz, 325 MHz, considering 3-$\sigma$ upper limits and observing time. The diagonal-filled spaces in these figures symbolise the allowed parameter space, whereas white space indicates the forbidden space. Dark magenta, dark slate grey, and maroon colours represent the allowed parameter space for band 5, band 4, and band 3 of GMRT, respectively. The vertical pink region indicates the 1-$\sigma$ span of the number density ranges from afterglow modelling. The blue dash-dot line represents the maximum rotational energy of a magnetar.}
\label{fig:parameter_space}
\end{figure*}

\begin{table}
\caption{Parameters from afterglow modelling}
\centering
\smallskip
\begin{tabular}{l c c c}
\hline \hline
GRB Name         & Ejecta Mass           	    & Number density        & References     \\
              &($M_{\odot}$)	 	    & ($\rm cm^{-3}$)                    &              \\
\hline 

050709    & 0.05, 0.1  &  $1.0_{-0.4}^{+0.5}$      & 1,2\\  

100625A   & -----  & $\leq$1.5     & 3 \\  

140903A   & 0.01   &   $3.40_{\rm -1.6}^{+2.9} \times 10^{-3}$  & 1,4 \\

160821B  & $\leq$ 0.006  & $0.13_{-0.04}^{+0.05}$    & 5,6 \\
\hline                            
\end{tabular}
\newline
\newline
\noindent
\noindent
\footnotesize{Note: $1$ - \citet{2015ApJ...815..102F}, $2$ - \citet{2016NatCo...712898J}, 3 - \citet{2013ApJ...769...56F}, 4 - \citet{2016ApJ...827..102T}, 5 - \citet{2019MNRAS.489.2104T}, 6 - \citet{2019ApJ...883...48L}}
\label{tab:parameter}      
\end{table}

In figure \ref{fig:parameter_space}, we present the allowed region of the $E_{\rm rot} - n_0$ plane for each burst. The maximum rotational energy of a magnetar $\sim 10^{53}$ erg \citep{2015MNRAS.454.3311M} is indicated by a horizontal line. The pink shaded region indicates the 1-$\sigma$ uncertainty of the number density of ambient medium available in the literature based on afterglow modelling as mentioned in Table \ref{tab:parameter}. As the afterglow data for most of these bursts are sparse, the inferred $n_0$ is subject to assumptions of downstream magnetic field (represented by $\epsilon^{\rm AG}_{B}$) of the afterglow fireball. However, a lower limit to the ambient density can still be obtained, assuming the highest possible $\epsilon^{\rm AG}_{B}$. Hence, the ambient density inferred from afterglow data still has the potential to constrain the rotational energy of the magnetar. 

The first panel of figure \ref{fig:parameter_space} depicts the parameter space plot of GRB 050709 in band 4 and band 3, where we can place a tight constraint on the rotational energy $E_{\rm rot} \sim (1.5 - 2) \times 10^{51}$ erg that is two orders lower than the maximum rotational energy expected from stable neutron star as given in \citet{2015MNRAS.454.3311M}. For GRB 100625A, the maximum energy possible by a magnetar is $E_{\rm rot} \sim 5 \times 10^{51}$ erg, which is one order of magnitude below the energy range given in afterglow study \citep{2013ApJ...769...56F}. The maximum rotational energy of GRB 160821B $E_{\rm rot}$ lies $\sim 0.7 \times 10^{52}$ erg, which is in agreement with the maximum energy of magnetar value given in \citet{2020ApJ...902...82S}. As GRB 160821B was observed before the predicted time span of the merger ejecta emission, late-time monitoring would lead to a much better constraint on the maximum rotational energy.

GRB 061210 is the only burst in our sample for which afterglow number density constraints are unavailable in the literature. Therefore, we cannot limit the energetics of a potential magnetar in this case. For GRB 140903A, the number density strip intersects with the permissible parameter space between $(0.6 - 2) \times 10^{52}$ erg for both frequency bands, which indicates its maximum possible energy. After considering the fixed values of $M_{\rm ej}$, $\epsilon_B$ and other canonical parameters, none of the GRBs we analysed is consistent with a magnetar central engine with rotational energy of $\sim 10^{53}$ erg which corresponds to a maximally rotating magnetar.

\subsection{Radio emission without magnetar injection}

We consider the case where energy injection from a short-lived magnetar energises the kilonova ejecta to mildly relativistic velocities. This would not have happened if the merger resulted in a prompt black hole. Without additional energy injection, the ejecta will remain sub-relativistic, and its deceleration will further be delayed, leading to a late (several 10s of years for typical parameters) onset of the radio emission \citep{2011Natur.478...82N}. In addition, the radio source will be fainter. Therefore, the numbers one can obtain from assuming a prompt collapse are less constraining. 

Nevertheless, the shock dynamics remain the same, with the rotational energy being replaced by the original kinetic energy, $1/2 M_{\rm ej} v_{\rm ej}^2$ of the ejecta, where $v_{\rm ej}$ is the initial velocity of the merger ejecta. Therefore, the upper limits on the energy obtained from our model could be translated to an upper limit on the velocity of the ejecta for a black hole central engine model. For the GRBs we have considered here, this leads to $\beta_{ej}\leq 0.05 - 2.5$. This range is not very different from the velocity of the merger ejecta derived from the optical observations of the kilonova \citep{2017Natur.551...64A, 2017ApJ...848L..17C}.

However, it must be mentioned that the kilonova ejecta may have a velocity profile, unlike the single velocity shell we have considered here. The fast-moving ejecta carrying a lower mass can decelerate early enough (in the timescale of months to years) and produce bright X-ray/radio emissions. Several authors explored this model after GRB 170817A \citep{2019MNRAS.487.3914K, 2019ApJ...886L..17H, 2021ApJ...914L..20B} and constraints on the velocity profile emerged after the detection of an X-ray excess in GRB170817 \citep{2020MNRAS.498.5643T, 2022ApJ...927L..17H}.

\subsection{Kilonova ejecta detectability in low frequencies}
\label{kilonova_detectability}

We also see that the fireball turns optically thin for standard parameters before $t_{\rm dec}$. This implies that the expected flux density $f_{\nu}$ is proportional to $\nu^{-0.6}$ (for a $p=2.2$) for almost all late-time radio observations done so far; hence $MHz$ frequencies can provide better constraints than GHz frequencies from equally deep kilonova remnant searches. Upper limits in $600$~MHz and $6$~GHz differing by a factor of $3$ can provide similar constraints on the kilonova ejecta and magnetar energy. Because of this, though uGMRT limits are relatively shallower compared to those from VLA and ATCA observations in the literature, we arrived at equally significant constraints from our study. For the maximum rotational energy of potential magnetars, \citet{2016ApJ...831..141F} obtained a range of $2 \times 10^{51} - 5 \times 10^{54}$~erg in a sample of 9 GRBs with VLA upper limits of $\sim 20 \mu$Jy at timescales of $1.2 - 7.7$~years. For a sample of 9 bursts observed by VLA, \citet{2020ApJ...902...82S} constrained the maximum energy of the magnetar to be (0.6 - 17.6) $\times 10^{52}$ erg. Another sample of 17 bursts observed by VLA and ATCA, \citet{2021MNRAS.500.1708R} obtained constraints in the rotational energy as (2 - 5) $\times 10^{52}$ erg. Two of the bursts in our sample are common with \citet{2020ApJ...902...82S} while three are common with \citet{2021MNRAS.500.1708R}. The maximum energy values of a magnetar (0.05 - 2) $\times 10^{52}$ erg, obtained in our study with GMRT/uGMRT data, are consistent with the values inferred by these authors.

Motivated by this, we explored the detectability of kilonova radio emission for uGMRT band-3 ($400$~MHz). We find that even for the most desirable parameters such as a maximally rotating magnetar ($E=10^{53}$~erg), low ejecta mass ($M_{\rm ej} = 0.001M_{\odot}$), and high ambient density ($n_0=0.5$), a flux density above $30 \mu{\rm Jy}$ is not possible except for nearby ($z<0.06$) bursts. Therefore, kilonova searches have better chances for detection if attempted for nearby bursts having potentially denser ambient medium inferred from afterglow light curves.

\section{Summary}
\label{summary}

We conducted a comprehensive study on the late-time merger ejecta emission of short GRBs using low-frequency radio wavelengths. As proposed by \citet{2014MNRAS.437.1821M}, for a magnetar central engine, the ejecta gets re-energised, and the interaction of this ejecta and the surrounding ambient medium can produce a delayed emission, which is expected to peak in the low-frequency radio regime. We, therefore, started a search program with the legacy GMRT/uGMRT in low frequencies to detect the merger ejecta emission in nearby short GRBs and thereby probe the existence of a stable magnetar remnant. Our sample comprises five bursts within z $\leq$ 0.5 that exhibit complex features in the X-ray light curves. The observations were carried out at 610 and 325 MHz frequencies of legacy GMRT and band 5 and band 4 of uGMRT between $2 - 11$ years since the burst trigger time. 

We found no late-time radio emission near the \textit{Swift} XRT positions of the GRBs in our sample. Compared to the previous studies, a few improvements are implemented in the observational strategy and the modeling. While the other studies focused on the higher frequency bands of VLA and ATCA, we observed with the low-frequency bands of GMRT as the expected spectrum of the late-time merger ejecta emission peaks around 600 MHz. The observations performed with a telescope like GMRT, which is sensitive to low frequencies, make this an important study. We obtained the dynamics of the decelerating ejecta using the generic model developed by \cite{PeerDynamcis2012}, which does not assume either non-relativistic or ultra-relativistic initial velocity. The ejecta dynamics were not considered in the earlier studies by \citet{2014MNRAS.437.1821M, 2016ApJ...819L..22H,2016ApJ...831..141F}. However, our study does not consider the effects of magnetar before the deceleration epoch, and the radio flares originated from the jet preceding the ejecta into the ambient medium.

In calculating the model flux, we have assumed equipartition between electrons and magnetic field. A simplified ejecta structure of a single shell without velocity stratification was considered. We ignored the initial phase, where the ejecta is accelerated to the velocity corresponding to the magnetar rotational energy. The impacts due to the jet in the ambient medium have been ignored.

Our model light curves show that the variation in all the model parameters significantly impacts the light curves. Comparing the 3$\sigma$ upper limits with the theoretical light curves, we placed constraints on the $E_{\rm rot}-n_0$ plane, as these two are the most decisive parameters in our model. Along with the inferred ambient medium density from the afterglow, this method allows us to place limits on the maximum rotational energy of the potential magnetar at $E_{ \rm rot} \leq (0.05 - 2)\times 10^{52}$ erg. We can exclude the magnetar central engine with $10^{53}$ erg for all the bursts as the highest possible rotational energy allowed by the radio upper limits lies much below $10^{53}$ erg. As no inferences on $n_0$ are available for GRB 061210 from afterglow literature, the constraints for this burst suffer from large uncertainties. If we consider the maximum energy of the magnetar to be $10^{52}$ erg, only GRB 050709 will be discarded from the probability of having a magnetar central engine. Despite having extended X-ray emission in GRB 050709, which is attributed to the presence of magnetar, it has the lowest maximum allowed energy. This can happen if the equation of state of the neutron star remnant is very soft. The absence of a stable neutron star as the remnant for the significant fraction of short GRBs states that the binary NS merger may directly collapse to BH \citep{2017ApJ...844L..19P},  which indicates a softer equation of state or a different merger scenario like NS - BH merger \citep{1999ApJ...527L..39J, 2008MNRAS.385L..10T, 2020ApJ...895...58G}. 

The late-time merger ejecta emission in short GRBs is a unique tool to get insights into the progenitor system. It can complement the afterglow and kilonovae studies. GRB 170817A is the most favourable object to identify the signature of merger ejecta emission due to its proximity ($\sim$ 40 Mpc) compared to other short bursts. Using our model, the estimated deceleration time of the kilonova ejecta in GW 170817 is beyond the last observation time of the source presented in \citet{2022ApJ...927L..17H} even for the most promising parameters. In the near future, it may be possible to detect merger ejecta emission from nearby short GRBs with the next-generation radio telescopes and prove the existence of a magnetar central engine. The upcoming Square Kilometer Array (SKA; \citealt{2021MNRAS.500.3821B}), in MHz frequencies, with increased sensitivity of $\mu$Jy level, will push the detection limits of merger ejecta emission at late times.

\section*{Acknowledgements}
The authors thank the referee for providing critical comments on the manuscript, which has improved the presentation of the results. We thank the staff of the GMRT that made these observations possible. GMRT is run by the National Centre for Radio Astrophysics (NCRA) of the Tata Institute of Fundamental Research (TIFR). AG thanks Ishwara-Chandra C. H. for kindly making the GMRT data analysis pipeline available. LR and KM acknowledge support from the grant EMR/2016/007127 from the Dept. of Science and Technology, India. K.G.A.~acknowledges support from the Department of Science and Technology and Science and Engineering Research Board (SERB) of India via the following grants: Swarnajayanti Fellowship Grant DST/SJF/PSA-01/2017-18, Core Research Grant CRG/2021/004565, and MATRICS grant (Mathematical Research Impact Centric Support) MTR/2020/000177.

\section*{Data Availability}

The radio data underlying this article is available in the article.



\bibliography{refag}{} 
\bibliographystyle{mnras}



\newpage
\appendix

\bsp	
\label{lastpage}
\end{document}